\documentclass[%
twocolumn,
amsmath,
amssymb,
aps,
prb,
caption=justified
]{revtex4-2}

\usepackage{float}
\usepackage[version=4]{mhchem}
\usepackage{acronym}
\usepackage{siunitx}
\usepackage{makecell}
\usepackage[skip=10pt]{subcaption}
\usepackage{lipsum}
\usepackage{xcolor}
\usepackage{graphicx}
\usepackage{dcolumn}
\usepackage{bm}
\usepackage{hyperref}
\usepackage{cleveref}
\usepackage{color,soul}
\usepackage{ragged2e}

\usepackage{xpatch}

\makeatletter
\xpatchcmd{\@ssect@ltx}{\@xsect}{\protected@edef\@currentlabelname{#8}\@xsect}{}{}
\xpatchcmd{\@sect@ltx}{\@xsect}{\protected@edef\@currentlabelname{#8}\@xsect}{}{}
\makeatother

\emergencystretch=\maxdimen
\begin{document}

\title{Comprehensive Structure Exploration and Thermodynamics of Graphitic Heteroatom Doped Graphene Superstructures}

\author{Benedict Saunders}%
\affiliation{Department of Chemistry, University of Warwick, Gibbet Hill Road, CV4 7AL, United Kingdom}
\author{Lukas H{\"o}rmann}%
 \email{lukas.hoermann@univie.ac.at}
\affiliation{Department of Chemistry, University of Warwick, Gibbet Hill Road, CV4 7AL, United Kingdom}
\affiliation{Department of Physics, University of Warwick, Gibbet Hill Road, CV4 7AL, United Kingdom}
\affiliation{University of Vienna, Faculty of Physics, Kolingasse 14-16, Vienna, Austria}
\author{Reinhard J. Maurer}%
\email{reinhard.maurer@univie.ac.at}
\affiliation{Department of Chemistry, University of Warwick, Gibbet Hill Road, CV4 7AL, United Kingdom}
\affiliation{Department of Physics, University of Warwick, Gibbet Hill Road, CV4 7AL, United Kingdom}
\affiliation{University of Vienna, Faculty of Physics, Kolingasse 14-16, Vienna, Austria}

\date{\today}

\begin{abstract}
Graphene has been studied in detail due to its mechanical, electrical, and thermal properties. It is well documented that the introduction of dopants or defects in the lattice can be used to tune material properties for a specific application, such as in electronics, sensors, or catalysis. To design graphene with specific properties, one must achieve control over the composition and concentration of defects. This requires a fundamental understanding of the stability of defects and their interaction in a superstructure. We present a comprehensive defect structure determination approach that enables close to exhaustive enumeration of all relevant defect structures. The approach uses a combination of Density Functional Theory and machine learning to build a transferable energy model for defect formation. We show the capabilities of our approach for free-standing graphene with heteroatom defects, establishing a thermodynamic model to investigate how temperature affects the configuration space of doped graphene. Our analysis yields physical insights into defect interactions. A characteristic peak in the heat capacity indicates a structural transition. We provide a mechanistic explanation for this behavior and demonstrate how the transition temperature shifts with different thermodynamic reservoirs.
\end{abstract}

\maketitle

\section{Introduction}

Since its isolation in 2004,\cite{novoselov2004electric} graphene has garnered significant interest owing to its remarkable electronic,\cite{novoselov2005two, nato2009electronic} thermal,\cite{balandin2011thermal} and mechanical properties.\cite{tsoukleri2009subjecting}
Pristine, undoped, free-standing graphene is a zero-gap semiconductor, evidenced by its characteristic Dirac cones at the $\Gamma$ point. 
A significant amount of research over the past two decades has been dedicated to modifying the structure of graphene to include various dopant atoms or molecules, and even vacancies, with the aim of effectively tuning the chemistry of the material to achieve a particular function.\cite{mazanek2019ultrapure, wang2020will, han2019defective} 
Altering the composition or topology of graphene can be used as a means to tune its band structure.\cite{rizzo2018topological, yankowitz2018dynamic} 
Doping with heteroatoms has been shown to open the bandgap, prompting extensive research on how various dopant configurations affect the band structure.\cite{ranidesigning2013}

\par Due to their semiconducting properties, graphene systems doped with heteroatoms have been studied meticulously for their potential applications as photocatalysts, \cite{zeng2016layer, bie2021design, chen2023implanting}  gas sensors,\cite{mirzaei2023n, nath2023pyridinic} as well as hosts for single-atom catalysts (SAC)\cite{he2020atomically, baby2021single, yang2022theoretical, alvarado2024influence}. 
Nitrogen-doped graphene (NDG), in particular, graphitic nitrogen, where a carbon atom is substituted for a nitrogen atom, is well studied; however, to achieve control over the doping level and dopant dispersity during growth requires understanding of the energetics and thermodynamics of the system in addition to the growth mechanism in the presence of the doping agent.
A significant problem that arises when modelling such systems is the sheer number of possible configurations that arise from doping graphene with even only a handful of nitrogen atoms. 
Various ways of experimentally growing NDG have been explored, including precursor molecules,\cite{usachov2011nitrogen} segregation from embedded carbon and nitrogen sources,\cite{zhang2011synthesis} and sputtering.\cite{zhao2012production}
Theory-based studies have investigated the carbon-nitrogen phase diagram, probing the limits of sustainable doping,\cite{shi2015much} identifying stable NDG structures,\cite{xiang2012ordered} and demonstrating how the choice of nitrogen and carbon sources can enable the growth of high-concentration NDG.\cite{bu2020design}
Methods such as cluster expansion, genetic algorithms,\cite{shi2015much} or particle swarm optimisations\cite{xiang2012ordered, bu2020design} enabled the exploration of dopant interactions and the material’s potential energy surface.
However, these approaches are mainly focused on the thermodynamic ground state, seeking the lowest-energy configuration at each composition.
The focus on only the lowest energy configuration is insufficient for investigating the thermodynamic properties of a system, which are determined by an ensemble of possible configurations, and not just the lowest-energy state.
Nested sampling offers a promising route to access thermodynamic properties,\cite{partay2014nested, yang2025freebird} but it is less suited to crystalline materials, where large regions of configuration space are sterically inaccessible. 
Especially in large-scale defect superstructures, the vast combinatorial complexity remains a major limiting factor.

\par Focusing on NDG, here we resolve this combinatorial challenge by leveraging a recent machine-learning-assisted structure search method, SAMPLE,\cite{hormann2019sample} and hence extract thermodynamic information about the system through the application of statistical mechanics.
SAMPLE utilises Bayesian regression to construct an efficient model for predicting the formation energy as a sum of on-site and pairwise interaction contributions. The final model is highly efficient, enabling brute-force structure search through combinatorial enumeration of millions of defect structures.
SAMPLE was originally developed to investigate the configuration space of adsorbed molecules on flat surfaces. To efficiently explore the vast configurational and compositional landscapes of NDG, we have modified the SAMPLE code, extending its capability to predict the formation energies of defective 2D materials.
By considering all possible configurations and their energies as individual states within a statistical mechanical framework, we gain insight into the thermodynamic behaviour of the material and enable the calculation of temperature-dependent properties, such as the heat capacity, which in turn can be used to predict the phase stability.

\section{Methods}

\subsection{System Setup and Assumptions}

Our thermodynamic analysis is an approximation of a chemical vapour deposition (CVD) experiment.
The CVD growth process occurs at elevated temperatures (\SI{1000}-\SI{1300}{\kelvin}) and doped configurations are thermodynamically and kinetically accessible,  permitting exploration of the rich configurational space. Once the growth process is complete, the system is slowly cooled to a given temperature, allowing it to settle into a distribution of structures given by our thermodynamic predictions. Our calculated heat capacity, therefore, does not correspond to a low-temperature dynamic transition but rather provides a thermodynamic descriptor of the relative populations of different NDG configurations created upon cooling. NDG can exhibit two primary classes of defects: substitutional defects, where a nitrogen atom replaces a carbon atom, and intrinsic lattice defects, which are structural flaws in the carbon lattice.\cite{fujimoto_formation_2011, kumar2012thermal, doustkhah2023templated} Graphitic nitrogen is experimentally known to be the dominant nitrogen species under high-temperature CVD growth and post-annealing conditions, where substitutional nitrogen becomes the most abundant configuration.\cite{lv_nitrogendoped_2012, trusovas2013reduction, doustkhah2023templated} Motivated by this, the main manuscript focuses on graphitic nitrogen–doped graphene. However, SAMPLE itself is agnostic to the specific defect type. To illustrate this generality, Section S3.6 of the ESI includes compositional phase diagrams generated with a SAMPLE model for tri-pyridinic nitrogen defects.

\subsection{Formation Energies Calculations}

First-principles electronic structure calculations are performed using density functional theory (DFT) to generate training sets, validation sets, and reference potentials. All DFT calculations were performed using the all-electron, atom-centered numerical orbital code FHI-aims\cite{blum2009ab, havu2009efficient} at the PBE\cite{perdew1996generalized} level of theory, with the addition of the Tkatchenko-Scheffler Van der Waals correction method.\cite{tkatchenko2009accurate} Graphene is described as a free-standing monolayer with a perpendicular vacuum slab of \SI{15.0}{\angstrom}, which was determined to be sufficient based on convergence testing (see Sections S1 and S2.2 of the ESI for convergence tests). Custom k-grids, based on a generalised Monkhorst-Pack scheme,\cite{monkhorst1976special} were used to ensure uniform k-grid sampling across different unit cell shapes and sizes. Further details are provided in Section S1 of the ESI.

All DFT calculations were conducted using a \textit{tight} default basis set. The energy tolerance was set to $1.0\times10^{-6}$ eV for electronic convergence, and the geometry optimisations were performed using a force tolerance of \SI{0.01}{\electronvolt\per\angstrom}. The angles of the graphene unit cell were fixed during geometry optimisations.  All initial graphene geometries were generated from the experimentally determined unit cell.\cite{meyer_structure_2007}

\subsection{SAMPLE for Defects in 2D Materials}

\begin{figure}
    \centering
    \includegraphics[width=\linewidth]{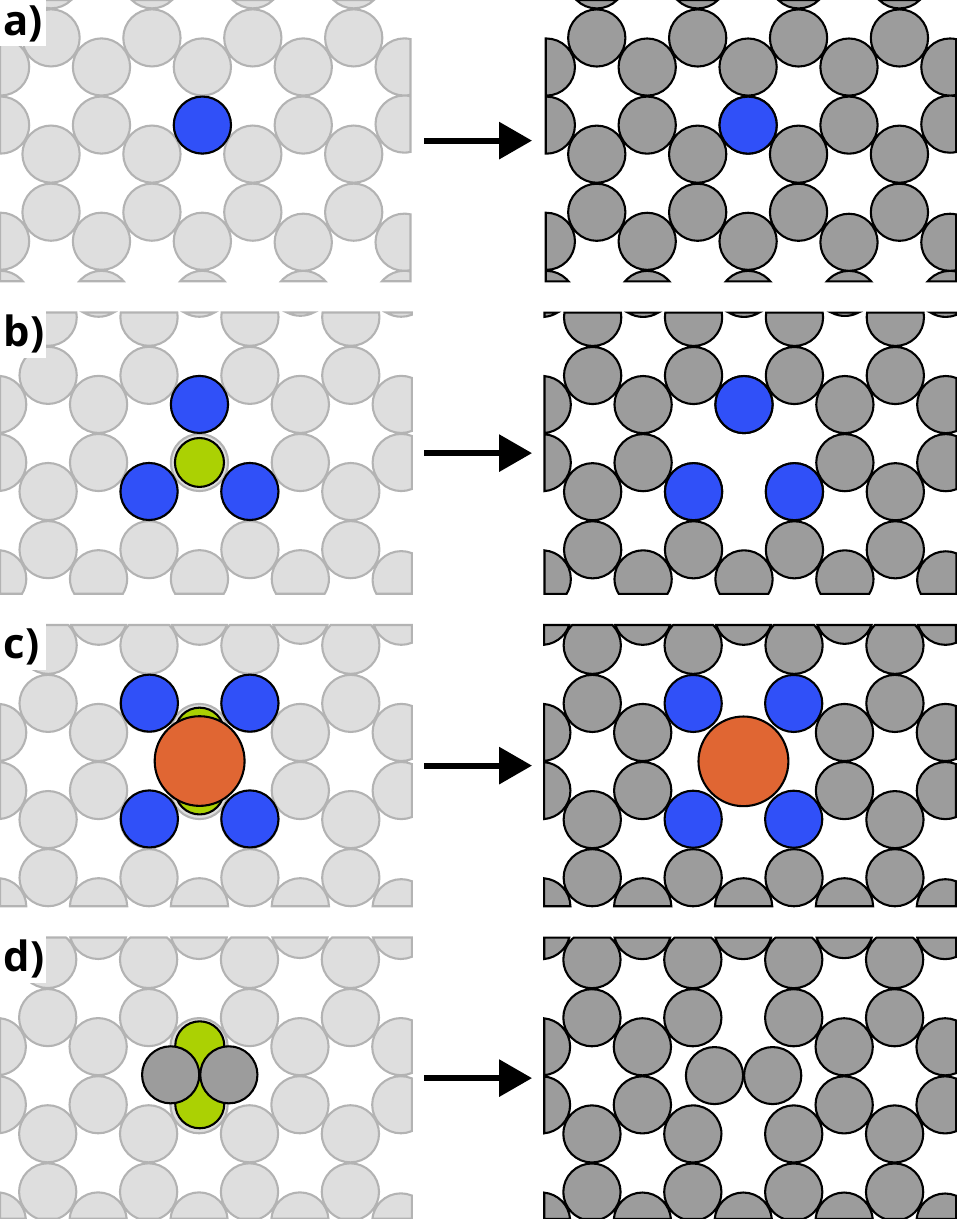}

    \caption{Various defect templates for graphene (first column) and the respective defective structure created with the template (second column): (a) Graphitic nitrogen; (b) Pyridinic nitrogen; (c) a \ce{FeN$_4$} metal complex; (d) a Stone-Wales defect; In (b), (c), and (d) the Rm atom type (yellow-green) is added to indicate the removal of an atom to form a vacancy.}
    \label{fig:defects_geometries}
\end{figure}

A crucial step for investigating the thermodynamic properties of defective 2D materials is the generation of a comprehensive set of possible defect superstructures.
SAMPLE tackles the combinatorial challenge posed by surface defects by restricting the degrees of freedom that define the search space, achieved through a process of coarse-graining, which approximates the continuous search space by treating adsorption sites as positions on a grid with discrete translations and rotations of the sites.
With such an approximation, the search space of a fixed substrate supercell with a fixed number of adsorbates becomes finite.
This is enforced by requiring the substrate supercell and the adsorbate unit cell to be commensurate. 
With this, the vectors that describe the netplane of the substrate $\mathbf{s}$ and the adsorbate, or defect in our case, $\mathbf{d}$ can be mapped onto one another with
\begin{equation}
\begin{pmatrix}
    \mathbf{d}_1 \\ \mathbf{d}_2
\end{pmatrix} = \mathbf{M}\cdot\begin{pmatrix}
    \mathbf{s}_1 \\ \mathbf{s}_2
\end{pmatrix}
\end{equation}
where all elements of the $2\times2$ epitaxial matrix $\mathbf{M}$ must be integers, and $\mathrm{det}(\mathbf{M})\neq0$.\cite{hermann2017crystallography} By defining a substrate supercell and a concentration of defects, we can enumerate all possible structures for that combination and keep those that are symmetrically unique. Just as we can define an adsorbate geometry commensurate with the substrate supercell, we can define a defect geometry commensurate with the graphene supercell. In the case of a single graphitic nitrogen atom, the unit cell used to describe the defect template is equivalent to that of pristine graphene (\cref{fig:defects_geometries}a and \ref{fig:defects_geometries}b).
By approximating surface-adsorbate geometries as large-scale commensurate structures, SAMPLE is well-suited to model systems such as doped graphene, due to its highly regular lattice.

SAMPLE stores geometric information in a configuration hash, a short tuple of integer numbers, representing the epitaxial matrix of the supercell, the positions, and types of defects. The positions are given by a unique number scheme of the possible defect sites in the substrate lattice. The defect types are indices pointing to a lookup table. This table stores the optimised geometric structures of the defects (in isolation), called defect templates. The configuration hash has a one-to-one correspondence to the structure and is invariant under symmetry operations applied to the configuration.

To enable the automated generation of 2D-defect superstructures, we made two modifications to the SAMPLE code. The first enables the substitution of atoms in the substrate, such as one would need for a graphitic defect - any atoms in the substrate geometry that overlap with an atom in the defect template, within some tolerance, are automatically overwritten. The second modification is the introduction of a new atomic species that can be included in the defect geometry: removium, designated `Rm'. As the name suggests, it deletes any atom at the location of the `Rm`, but otherwise behaves the same as other atoms (e.g., adhering to symmetry operations). As such, `Rm' can effectively create vacancies in the substrate lattice, but can also be used to delete atoms to create more complex defects too, such as FeN\textsubscript{4},\cite{jeong_co_2025} or the Stone-Wales defect (Figures~\ref{fig:defects_geometries}c and~\ref{fig:defects_geometries}d).

When introducing a defect into a lattice (or an adsorbate onto a surface), the SAMPLE approach assumes that the substrate lattice remains unchanged, and all lattice atoms (apart from ones removed or replaced) retain their original positions in the undoped substrate supercell. Typically, the inclusion of defects results in rearrangements in the lattice, but this is neglected by SAMPLE, resulting in an overestimation of formation energies. The lattice distortion that occurs when relaxing a single heteroatom-defect-structure is small (see \nameref{sec:defect_defect_interactions} and Sections S2.2 and S2.3 of the ESI). This is consistent with previous reports: substitutional graphitic nitrogen slightly reduces the lattice constant,\cite{ullah2017rectangular} while recent studies reported sublattice symmetry breaking and C-C bond-length variation with dopant number, although they do not observe large global shear or angular deviations.\cite{yutomo_effect_2021, dimov2023tailoring} While high defect concentrations can more substantially alter the structure and stability of NDG, these configurations have a negligible impact on the thermodynamic properties, as their Boltzmann factors are small. Therefore, we expect that the assumption of a fixed unit cell does not significantly affect the thermodynamics of graphitic-doped NDG. However, some defects, such as the FeN\textsubscript{4} or tri-pyridinic substructures shown in Figure \ref{fig:defects_geometries}, are expected to induce significant lattice distortions. For this reason, we have elected to focus primarily on graphitic nitrogen defects in this study. In section S3 of the Supplementary Material, we also demonstrate the modelling of tri-pyridinic nitrogen defects with this approach, albeit only for a restricted region of the configurational space.

\subsection{Learning the Formation Energies}

SAMPLE leverages Bayesian regression for accelerated evaluation of formation energies $E_\text{i}$ per defect. For consistency with later chapters, we introduce the index i, which denotes the $i^\text{th}$ configuration. For graphitic nitrogen defects in graphene, a carbon atom is replaced by a nitrogen atom. We assume the system is in contact with reservoirs for carbon and nitrogen, therefore representing the system in a grand canonical ensemble. The energy required to exchange species with the reservoir is given by the respective chemical potentials. While the SAMPLE approach is agnostic to the choice of chemical potential, the choice affects the predicted thermodynamic behaviour (see figure \ref{fig:thermodynamics}). To demonstrate this, we select different chemical potentials:
\\

\noindent
\underline{Carbon chemical potential ($\mu_\text{C}$)}
\begin{itemize}
    \item \textbf{Graphite}: The binding energy of a carbon atom in bulk graphite is a standard literature reference for carbon systems.\cite{khan2018binary, bu2020design, yang2015structural}
    \item \textbf{Methane}: The binding energy of carbon in methane is used to reflect CVD growth of graphene, which typically utilizes small hydrocarbon precursors.\cite{wang2012review}
\end{itemize}

\noindent
\underline{Nitrogen chemical potential ($\mu_\text{N}$)}
\begin{itemize}
    \item \textbf{Dinitrogen}: The binding energy of a nitrogen atom in molecular nitrogen is the conventional reference,\cite{pham_how_2021, yutomo2021effect} though its extremely high binding energy often misrepresents reactive environments.
    \item \textbf{Ammonia}, \textbf{pyridine}: The binding energy of nitrogen in ammonia or pyridine is representative of the reactive nitrogen-containing molecules commonly supplied in CVD experiments.\cite{wang2012review} \end{itemize}

In all cases, where the chemical potentials used are not explicitly stated, dinitrogen-graphite is used. Details on the exact calculations of the chemical potentials are given in Section S2.1 of the ESI. The exact values of the chemical potentials, $\mu_C$ and $\mu_N$,  used in eq. \ref{eq:formation_energy} are given in Table S1. The formation energy can be written as follows:
\begin{equation}
    E_{\text{i}} = \frac{1}{N_{\text{i}}} \cdot \left( E_{\text{SYS}} - E_{\text{GR}} + N_{\text{C}} \cdot \mu_{\text{C}} - N_{\text{N}} \cdot \mu_{\text{N}} \right)
    \label{eq:formation_energy}
\end{equation}

Here $E_{\text{SYS}}$ and $E_{\text{GR}}$ are the energies of the defective and pristine graphene sheet, respectively. $N_{\text{i}}$ is the number of defects within the supercell adopted in the computation, while $N_{\text{N}}$ and $N_{\text{C}}$ are the numbers of nitrogen atoms and carbon vacancies associated with these defects. To learn and predict the formation energy per defect, $E_{\text{i}}$, SAMPLE uses a cluster-expansion energy model with one-body energy interactions $g$ and two-body interaction energies between pairs of defects, $p$. 
\begin{equation}
E_{\text{i}}=\sum_\text{g} n_\text{g} U_\text{g} + \sum_\text{p} n_\text{p} V_\text{p}
\label{eq:energy_model}
\end{equation}

Here $n_\text{g}$ and $n_\text{p}$ specify the occurrences of each of the different one- and two-body interactions within a given configuration.
The values in the vector  $\mathbf{U}$ correspond to the dopant-lattice interaction in isolation, analogous to the adsorption energy of a single molecule on a surface, and $\mathbf{V}$ discretises the interaction energies between pairs of defects at different distances. Once trained, these vectors can be considered as the formation energies of isolated defects and the energies of pairs of defects on the coarse-grained lattice, respectively.
Concatenating $\mathbf{U}$ and $\mathbf{V}$ into a single interaction vector $\boldsymbol{\omega}$, together with the model vector $\mathbf{n}$, which lists the occurrences of individual interactions in a configuration, the energy per dopant of a configuration can be evaluated using linear regression:
\begin{equation}
E_\text{i} = \mathbf{n}_\text{i} \cdot \boldsymbol{\omega}
\end{equation}

As $\mathbf{U}$ and $\mathbf{V}$, hence $\boldsymbol{\omega}$, contain the energy contributions of every possible combination of one- and two-body terms considered by a given SAMPLE model, only $\mathbf{n}$ changes with configuration. As such, all of the configurations to be considered can be agglomerated into a single model matrix $\mathbf{X}$, such that the energies of a configuration set can be simply expressed as
\begin{equation}
\mathbf{E} = \mathbf{X} \cdot \boldsymbol{\omega}
\end{equation}

SAMPLE is trained using Bayesian regression to fit the vector $\omega$. The reason for using this method is twofold: Firstly, fitting $\omega$ is an underdetermined problem, and there are far more interactions in the configuration space than one could reasonably have in the training data. This means that we can fit the model on only a few hundred DFT datapoints. Secondly, 
\begin{equation}\label{eq:bayes}
p(\boldsymbol{\omega}|\boldsymbol{E}_\text{DFT})=\frac{p(\boldsymbol{E}_\text{DFT}|\boldsymbol{\omega})p(\boldsymbol{\omega})}{p(\boldsymbol{E}_\text{DFT})}
\end{equation}

Bayes' theorem, given in Equation \ref{eq:bayes}, allows SAMPLE to ascertain the expectation values and respective probability distributions of the one- and two-body vectors.
In SAMPLE's case, the denominator $p(\boldsymbol{E}_\text{DFT})$, termed the \textit{marginal probability}, can be neglected since it is simply a normalisation constant.
The \textit{prior}, $p(\boldsymbol{\omega})$, and the \textit{likelihood}, $p(\boldsymbol{E}_\text{DFT}|\boldsymbol{\omega})p(\boldsymbol{\omega})$ are both Gaussian distributions, and therefore their product equals the \textit{posterior}, $p(\boldsymbol{\omega}|\boldsymbol{E}_\text{DFT})$.
The likelihood determines the probability that a given interaction vector $\boldsymbol{\omega}$ will reproduce the energies $\boldsymbol{E}_\text{DFT}$ calculated with DFT for a given set of configurations (i.e., the training data) within a predefined tolerance.
This training set consists of a few hundred DFT calculations of defect superstructures that are sampled from the comprehensive configuration space discussed in the previous section, through d-optimal selection.\cite{pukelsheim2006optimal}
The prior, on the other hand, provides the physical knowledge about the system, and as such, we can bias the likelihood such that the model may give more accurate predictions, typically using data from DFT calculations.
A more thorough explanation of the SAMPLE method can be found in the original paper.\cite{hormann2019sample}

\subsection{Thermodynamics with Formation Energies}

The ability of SAMPLE to comprehensively map the configuration space enables brute-force calculation of the partition function and all thermodynamic properties that can be derived from it. For each size of the simulation cell (see section S3.1 of the ESI), the composition space can be treated as a grand canonical ensemble for fixed values of $\mu_C$ and $\mu_N$. In our thermodynamic formulation, each microstate corresponds to a specific NDG configuration. To make systems with different coverages comparable, we determine the energy per simulation cell:
\begin{equation}
    \epsilon_\text{i} = E_\text{i} N_\text{i}
    \label{eq:partition_function}
\end{equation}

Using the formation energy of the simulation cell, we construct the grand canonical partition function for each cell size. Hereby, we make use of the fact that the formation energy per atom already includes the chemical potentials that govern particle exchange between the system and reservoirs (see Equation \ref{eq:formation_energy}). This approach is appropriate because the simulation cell defines the fundamental periodic unit that accounts for the total energetic cost of a specific configuration, including many-body interactions and surface strain. By treating the entire cell as a single state in the ensemble, we ensure that the competition between different phases is governed by the total free energy of the interface, rather than being biased by the varying number of constituent atoms in different supercell geometries.
\begin{equation}
    \mathcal{Z} = \sum_\text{i} \exp\left(-\beta \epsilon_\text{i}\right)
\end{equation}

In this equation, $\beta = \frac{1}{k_BT}$, $k_B$ is the Boltzmann constant, and $T$ is the temperature of the system. Boltzmann averages of observables can be calculated directly. We can calculate the Boltzmann probability corresponding to each microstate $i$:
\begin{equation}
    P_\text{i} = \frac{\exp{(-\beta\epsilon_\text{i})}}{\mathcal{Z}}
\end{equation}
which we can then use to calculate expected formation energy and particle concentration (normed to intensive energy per lattice-site) as
\begin{align}
    \langle E_\text{form} \rangle &= \frac{1}{N_\text{tot}} \sum_\text{i} P_\text{i} \epsilon_\text{i}
\end{align}

However, when calculating the heat capacity in the grand canonical ensemble, fluctuations of defect concentration $\theta_\text{i} = \frac{N_\text{i}}{N_\text{tot}}$ need to be considered, where $N_\text{i}$ is the number of defects and $N_\text{tot}$ is the total number of atoms in the configuration. This is given by the variance $\text{Var}(\theta) = \langle\theta^2\rangle-\langle \theta\rangle^2$ of the concentration and the covariance between the energy and concentration $\text{Cov}(E, \theta) = \langle\theta E\rangle-\langle \theta\rangle\langle E\rangle$ (see Section S6 of the ESI for a detailed derivation). The heat capacity of a grand canonical ensemble can hence be expressed as follows:
\begin{equation}
    C_\text{V} = \frac{1}{k_BT^2}\left[\text{Var}(\epsilon)-\frac{\text{Cov}(\epsilon,\theta)^2}{\text{Var}(\theta)}\right]
\end{equation}
and hence the specific heat capacity is calculated trivially as
\begin{equation}
    c_\text{V} = \frac{C_\text{V}}{N_\text{tot}}.
\end{equation}

Applying these expressions, we can explore the thermodynamic behaviour of the system for a given surface area under fluctuating numbers of defects and hence gain further insight into the thermodynamic stability and the response to temperature changes.

\section{Results}
\label{sec:results}

The results of this study are divided into three parts. Firstly, we report an investigation into the interactions of defects in graphene, as predicted by DFT. The SAMPLE model is fitted based on these data and validated against the DFT reference data. Secondly, we apply our SAMPLE model to explore the vast configuration space of doped graphene, a task that would be computationally prohibitive based solely on DFT. Finally, the derivation of thermodynamic properties of graphene doped with graphitic-nitrogen environments provides insights into the stability and phase behaviour of the material under varying conditions.

\subsection{Defect-Defect Interactions}\label{sec:defect_defect_interactions}

When investigating the configuration space of a doped material, two main energy contributions must be considered: (i) the intrinsic formation energy of each defect in the host lattice, and (ii) the interaction energy between defects, which becomes particularly significant when defects are in close proximity. Both contributions play a crucial role in determining the overall formation energy and thus the underlying chemistry and physics of the system.

\begin{figure*}[t]
  \includegraphics[width=0.80\textwidth]{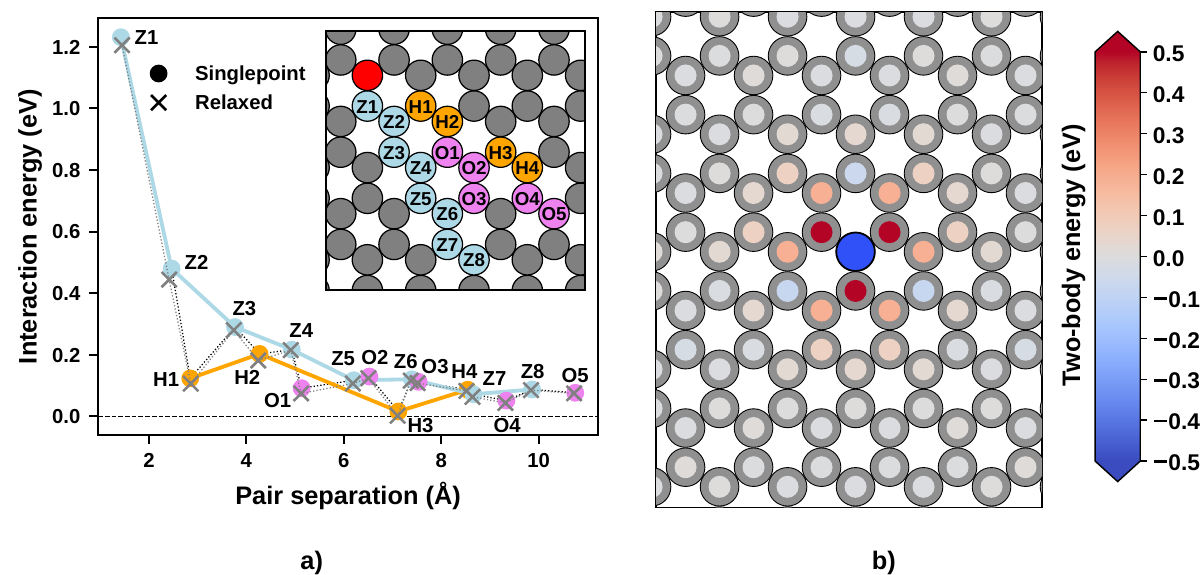}
  \caption{\justifying Nitrogen defect interaction energy with DFT (a) and two-body energies from SAMPLE model (b). The DFT interaction energies are split into three groups according to their relative position in the lattice: Z follows a zig-zag pattern, H `hops' over rings, and O configurations are combinations of both moves. The learned SAMPLE two-body energies in (b) are greater than \SI{0.50}{\electronvolt}; however, the colour scale has been shifted to emphasise that SAMPLE was able to infer the non-trivial interactions as observed with DFT.}
  \label{fig:interactions}
\end{figure*}

\paragraph*{\textbf{DFT calculations.}} 
The intrinsic formation energy $E_\text{intrinsic}$ is the energy required to exchange carbon for nitrogen in the graphene plane. We determine it using Equation \ref{eq:formation_energy} for a unit cell in which only a single defect is placed. The unit cell is chosen large enough that the defects no longer interact with each other (see Section S2.2 of the ESI). For the interaction energy, we can apply
\begin{equation}
    E_\text{interaction} = E_\text{pair} - 2 \cdot E_\text{intrinsic}
\end{equation}
where $E_\text{pair}$ is the formation energy (calculated with Equation \ref{eq:formation_energy}) of a structure containing a pair of defects.

We find that the formation energy of an isolated graphitic nitrogen site in graphene in the dilute limit is \SI{0.80}{\electronvolt}, calculated using a $12\times12$ supercell to ensure a safe margin beyond convergence (see Table S3). We also probed the effect of geometry optimisations on the formation energy by placing a single nitrogen defect into supercells of varying sizes. For all supercells larger than the primitive graphene unit cell, the total energy of the doped systems decreases between 0.03 and \SI{0.04}{\electronvolt}. Therefore, the lattice relaxation does not significantly affect the formation energies of isolated graphitic nitrogen defects and the interaction energy with periodic replicas of the defects. The dilute limit is reached with a $11\times11$ supercell, where the smallest distance between two nitrogen defects is \SI{14.8}{\angstrom} (see Figure S2 of the ESI).

With a term for the isolated formation energy, we can now calculate various interaction energies of graphitic nitrogen sites with DFT. Figure \ref{fig:interactions}a shows the interaction energy of multiple pairs of defects as a function of separation distance, covering 17 of the symmetrically unique interactions between \SI{1.4}{} and \SI{10.7}{\angstrom} in a $16\times16$ graphene supercell. Although there is a clear overall trend of decreasing interaction energy with increasing distance, the trend itself is very much non-trivial. Unsurprisingly, the largest interaction energy is exhibited by the configuration in which the two defects are on neighbouring lattice sites, \SI{1.20}{\electronvolt}.

The interaction profile between two dopand pairs in the lattice is non-trivial. To show this, we divide pair configurations into three categories: \emph{Z} configurations follow a zig-zag path, \emph{H} configurations effectively hop over consecutive rings, and \emph{O} configurations are some combination of the two (see Figure \ref{fig:interactions}a). The \emph{Z} configurations exhibit an interaction energy profile that closely follows an exponential decay with increasing distance. In contrast, the \emph{H} configurations include some of the most stable defect arrangements, particularly \emph{H1}, which represents a significant local energy minimum after hopping over a single hexagonal ring, and \emph{H3}, which corresponds to a second local minimum resembling a 4,4$^\prime$-bipyridine-like system. The \emph{O} configurations, which combine elements of both \emph{Z} and \emph{H} pathways, also show unexpectedly stable arrangements, particularly \emph{O1} and \emph{O4}. 

Additionally, Figure \ref{fig:interactions}a shows the gain in interaction energy upon geometry optimisation. This energy gain decreases with increasing pair distance and becomes negligible at larger separations that characterise energetically favourable structures. As detailed in Section 2.3 of the ESI, geometry relaxations can be neglected given our objective of investigating the qualitative thermodynamic behavior of NDG and given the fact that structure relaxation most strongly affects close distances with high energies that are less likely to contribute to low energy configurations.

\paragraph*{\textbf{SAMPLE.}}

Based on DFT reference data, a SAMPLE model was constructed to determine the formation energies of graphene supercells doped with graphitic nitrogen defects. Against the validation set, the model exhibits a root-mean-square-error of \SI{0.022}{\electronvolt} per defect for the formation energy prediction. Details regarding specific hyperparameters in the energy model can be found in Section S3.4 of the ESI. The two-body energies learned by SAMPLE are an approximation of the defect-defect interaction energies, and are illustrated as a heatmap in Figure \ref{fig:interactions}b. The two-body energies follow the trends observed in the DFT data in Figure \ref{fig:interactions}a, but are slightly offset from the DFT interaction energies. This is because the two-body energies and the DFT interaction energies have different references. As this offset is added to the one-body energies, this offset does not affect the total energies predicted with Equation \ref{eq:energy_model}. Further details regarding the model's accuracy are provided in Section S3.5 of the ESI.

\subsection{Configuration Space Exploration}

\begin{figure*}[htb]
  \includegraphics[width=0.80\textwidth]{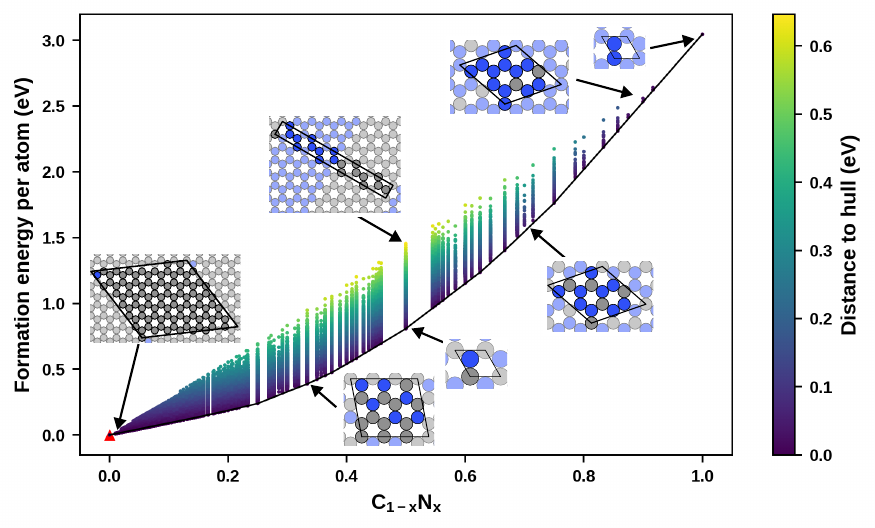}
  \caption{\justifying The compositional phase diagram of graphene doped with graphitic nitrogen sites. Pristine graphene is represented as a red triangle with a formation energy of \SI{0.00}{\electronvolt}. Also shown are a selection of structures to demonstrate the variety of supercell shapes and sizes considered by SAMPLE.}
  \label{fig:graphitic_hull}
\end{figure*}

To explore the configuration space of NDG, we generated a total of 162,239,485 configurations with a cell size of up to 100 graphene lattice sites. The nitrogen concentration ranges from \SI{1}{\percent} to \SI{100}{\percent}, with sampling biased toward low-nitrogen-content configurations (see Section S3.1 of the ESI), as these are energetically more favourable and are expected to dominate the thermodynamic properties (see \nameref{sec:results_themodynamics}). Symmetrically equivalent structures were screened out using a method described in Section S3.2 of the ESI, reducing the dataset to 105,614,115. Applying the SAMPLE model that we constructed above, we can illustrate the entire configuration space of graphitic nitrogen dopants by considering the formation energy as a function of nitrogen content, as in Figure \ref{fig:graphitic_hull}, where we use dinitrogen and graphite as the chemical potentials for nitrogen and carbon, respectively. We can describe the topology of the compositional phase diagram with a convex hull, and as such, the vertical distance from a given point to the hull is used as a measure of relative stability within the configuration space.

Overall, configurations with the fewest and most sparsely distributed dopants exhibit the lowest formation energies. However, the wide range of formation energies at each value of $\theta$ highlights that the specific atomic configuration, not just the defect concentration, plays a crucial role in determining the energetic favourability. For example, in Figure \ref{fig:graphitic_hull} at $\theta$~=~0.5, the most stable configuration exhibits an alternating arrangement, where the nitrogen and carbon atoms occupy graphene's inherent \emph{A} and \emph{B} sublattice sites, respectively, whereas the least stable configuration shows segregated nitrogen and carbon islands within the plane. It follows that the least stable configuration overall is \SI{100}{\percent} nitrogen content.

The formation energy (see Equation \ref{eq:formation_energy}) is a function of the chemical potentials for carbon and nitrogen, which depend on the experimental setup. While relative energies at a given composition are independent of the chemical potentials, the energy ranking of structures with different compositions may change. To gauge this effect, we present the potential phase diagrams based on the chemical potential combinations: pyridine-graphite, ammonia-graphite, ammonia-methane, in Section S5.1 of the ESI. Under these changes to the chemical potentials, the phase diagrams remain qualitatively similar, while the spread in formation energies and the slope of the lower convex hull change.

Finally, the gaps observed throughout the phase diagram in Figure \ref{fig:graphitic_hull} arise from the fact that we limit the configurations we consider to a maximum supercell size. Filling these gaps would require supercells with more than 100 graphene lattice sites, which would lead to a greater combinatorial explosion, becoming computationally intractable. Nevertheless, SAMPLE has generated a comprehensive set of configurations for the given constraints, which are used to evaluate the energy phase diagram of NDG deterministically.

\subsection{Thermodynamics}
\label{sec:results_themodynamics}

Equipped with a set of configurations and their corresponding formation energies, we can apply statistical mechanics to predict the thermodynamic properties of the system. Treating the entire configuration space as a grand canonical ensemble, we can construct the partition function explicitly by considering each configuration and its respective formation energy as a microstate of the ensemble (see Equation \ref{eq:partition_function}). To ensure valid sampling, our analysis uses only supercells with a fixed size of 100  graphene lattice sites. At this cell size, the thermodynamic properties are well converged, as shown in Figure S11.

\begin{figure}[h!]
    \centering
    \includegraphics[width=\linewidth]{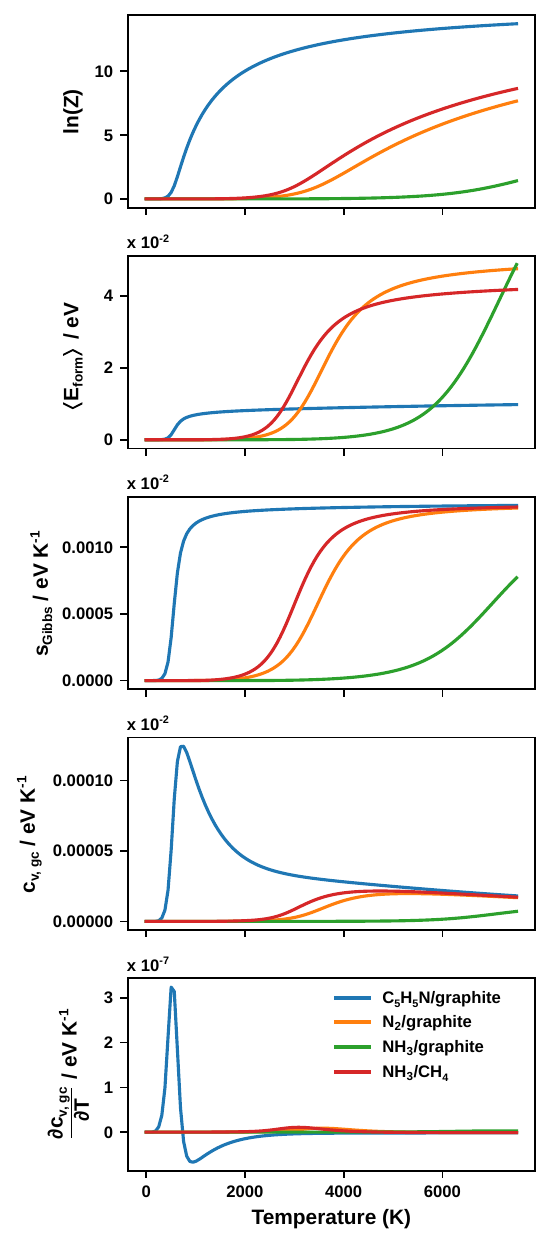}
    \caption{\justifying Predicted thermodynamic properties as a function of temperature, using all configurations with 100 atoms as a grand canonical ensemble. From top to bottom: the partition function, expected energy, entropy, heat capacity, and first derivative of the heat capacity calculated numerically.}
    \label{fig:thermodynamics}
\end{figure}

Figure \ref{fig:thermodynamics} shows the calculated thermodynamic quantities for NDG, illustrating how fluctuations in the concentration $\theta$ cause more and more states to become accessible with increasing temperature. This leads to a sharp increase in the internal energy and entropy. The temperature of this increase strongly depends on the choice of chemical potential. Irrespective of the chemical potential, the heat capacity exhibits an anomalous peak, which is typically indicative of a transition within the system. The position of this peak shifts with the chemical potential: For the pyridine-graphite chemical potentials, it lies at approximately \SI{700}{\kelvin}, for ammonia-methane at \SI{4700}{\kelvin}, for dinitrogen-graphite at \SI{5300}{\kelvin}, and for ammonia-graphite the peak is above \SI{7500}{\kelvin}. Temperatures of additional characteristic features in Figure \ref{fig:thermodynamics} are shown in Table S10. We note that temperatures of \SI{4000}{\kelvin} and more are above the sublimation point of graphite, where the model of NDG is no longer valid.

By taking the first derivative, we further show that this peak is continuous, with a clear minimum inflection point, indicating that the transition is complete at that temperature. Beyond this inflection point temperature, the derivative of the heat capacity becomes asymptotic to zero, indicating that the heat capacity itself has also become asymptotic, but to a non-zero value as a consequence of the transition.

Owing to the finite size of the systems we have investigated, we cannot achieve a defect concentration below \SI{1}{\percent}. In fact, there exists an infinite number of possible configurations, given an arbitrary supercell size. Sampling the configuration space uniformly with a maximum supercell size is expected to yield a representative, finite subset of NDG configurations. However, the sheer size of the configuration space forces us to focus on low-concentration configurations. However, due to the mostly repulsive defect interactions, low-concentration configurations dominate the thermodynamic properties in the present temperature range of \SI{0}{} to \SI{7500}{\kelvin}, as shown in Table S8. We expect that the addition of high-concentration configurations does not significantly affect the thermodynamic properties at lower temperatures: a more complete picture of the phase space would essentially push the critical point to a slightly higher temperature, while broadening the peak. A thorough discussion about sampling quality and the convergence of the thermodynamic quantities can be found in Sections S4.2 and S5.1 in the ESI.

\section{Discussion}

\paragraph*{\textbf{Defect-Defect Interactions.}}

Investigating the pair interaction energies with DFT yields similar results to those shown by previous studies.\cite{xiang2012ordered, bu2020design, houinterplay2012} Overall, nitrogen-nitrogen pair interactions are repulsive. This trend is likely a result of repulsive Coulombic electrostatic interaction, as previously argued.\cite{bu2020design} However, we find a minimum at the H3 position, as reported in the \nameref{sec:results}, corresponding to a configuration that does not appear to have been discussed in previous works. Nevertheless, the trends and anisotropies demonstrated by the defects' interaction profile are consistent with the literature - our results show that these profiles are more than just a function of distance and sublattice site, but very much dependent on the nature of the path taken between the defect sites. For example, the presence of a ``ring crossing'' between the two defects contributes to the stability of the configuration, while increasing separation via a zig-zag pathway just leads to a monotonic decay of the repulsion. Bu et. al. have previously argued that this is a result of the anisotropic distribution of the electron charges.\cite{bu2020design}

\paragraph*{\textbf{Compositional Phase Diagram.}}

We find that the compositional phase diagram is qualitatively robust against changes in chemical potential. Using the common literature references (dinitrogen and graphite), the topology of the phase diagram for graphitic-N defects in graphene matches that observed in previous works by Shi et al.\cite{shi2015much} and Bu et al.\cite{bu2020design}, the difference in this study being that we can guarantee that the configurations that lie on the convex hull are the most stable for a given dopant concentration. By investigating the configuration with much larger graphene supercells than previous works, we also demonstrate that regions of the phase diagram, which were previously shown to exhibit local concavities, are an artefact of the maximum size of system employed. 

The formation energy exhibits an overall exponential increase with increasing defect concentration, consistent with trends reported in the literature. In particular, the sharp change in the convex hull gradient at $\theta$~=~0.25 corresponds to the stoichiometry C\textsubscript{3}N and agrees with previous studies. No other comparably pronounced features are observed along the remainder of the convex hull. This is consistent with the hypothesis of Shi et al.\cite{shi2015much}, who proposed a `stability switchover' between $\theta$~=~0.33 and $\theta$~=~0.375, confirmed using phonon calculations in their study. However, the apparent range of this switchover may be an artefact of the small system sizes considered, while the sharp rise in formation energy at $\theta$~=~0.25 could mark the true onset of this effect.

\paragraph*{\textbf{Thermodynamic Properties and Phase Transitions.}}

We observe a transition in the heat capacity, evidenced by a peak. The position of this peak is strongly dependent on the choice of chemical potential. Using dinitrogen-graphite as chemical potentials, as commonly done in literature, we find this peak at a temperature of approximately \SI{5300}{\kelvin}, where basic assumptions of our model break down. This high temperature is the result of the strong binding energy of nitrogen molecules, which makes the incorporation of nitrogen into graphene energetically unfavourable. However, in CVD experiments, nitrogen is commonly supplied to the system in the form of nitrogen-containing molecules that give rise to a significantly lower effective chemical potential of nitrogen. Using pyridine-graphite as a source, for instance, shifts the peak to approximately \SI{700}{\kelvin}, which is in the range of common deposition temperatures.

This peak can be interpreted as a Schottky anomaly: At low temperatures, only a limited number of states are thermodynamically accessible. As the temperature surpasses a critical point, a significantly larger number of states becomes accessible, resulting in a peak in the heat capacity. As Figure \ref{fig:thermodynamics} demonstrates, there is a sudden increase in the system's entropy, which corresponds to the rise in the number of accessible configurations. Among the 105 million configurations explored, the compositional phase diagram of NDG exhibits two distinct regimes as dopant concentration increases: clustered, island-like defect superstructures and uniformly dispersed defects across the 2D plane. The most prominent example is illustrated in Figure \ref{fig:graphitic_hull} at \SI{50}{\percent} concentration; the configuration with the largest formation energy is composed of large strips of hexagonal 2D nitrogen and graphene, and the lowest energy configuration has C and N atoms at the \emph{A} and \emph{B} sublattice sites of graphene. It is impossible for a nitrogen-nitrogen neighbour defect pair not to be present in any given configuration with a defect concentration greater than \SI{50}{\percent}. Previous work\cite{chaban2015nitrogen} has shown that the presence of these nitrogen-nitrogen neighbour pair motifs is detrimental to the stability of the doped graphene structure. Naturally, clustered configurations are composed of large numbers of nitrogen-nitrogen neighbour pairs, which greatly increases their formation energy. Due to symmetry, low-energy configurations with maximally spaced nitrogen atoms are few in number compared to configurations containing clusters of nitrogen. Before the critical temperature is reached, only uniformly dispersed defect superstructures contribute to the partition function. Above the critical temperature, clusters of defect superstructures become available.

Similar peaks in heat capacities have been reported for various systems, such as order-disorder phase transitions in molecular solids,\cite{doi:10.1021/acs.jpclett.1c00289, szewczyk2021heat} charge disorder in bulk materials,\cite{PhysRevB.79.224111} binary clusters described with a Lennard-Jones potential,\cite{yang2025freebird} and for the Ising model.\cite{regeciova_localized_2025} Cairano et al.\cite{cairano_specific_2025} interpreted this peak as a third-order phase transition. Using the sine-Gordon model in a microcanonical ensemble on a square two-dimensional lattice, the authors found a ``roughness''-driven phase transition where the heat capacity displays a peak but remains a continuous function. In the sine-Gordon model, each site can assume arbitrary field values, qualitatively resembling our system, where nitrogen and carbon represent different field values when only one unit cell is considered. Both models feature one- and two-body interactions. However, the sine-Gordon model includes only next-nearest-neighbour interactions, while in the nitrogen-defect model, the most repulsive interactions occur at the closest range and decay rapidly (see Fig.~\ref{fig:interactions}). 

To study the effect of the energy distribution in the microstates, and how it may affect the energy fluctuations, and hence the 
heat capacity, we employ a toy model that allows us to arbitrarily generate energy spectra (see Section S5 of the ESI). We find that systems with degenerate ground states and final states result in a broader peak, reaching into very high temperatures. Conversely, for energy spectra with a single ground state, final state, but degenerate intermediate states, the heat capacity is much narrower. Such energy spectra are a result of the combinatorics governing the energetic favourability of structures: Only very few configurations of a system's degrees of freedom will lead to very low or high energies, while a large number of configurations have moderate energies. This was observed in several studies on surface structure prediction.\cite{hormann2019sample, doi:10.1021/acsnano.0c10065, PhysRevMaterials.2.043803, hormann2022bistable} This transition represents a combinatorially driven order-disorder phenomenon, where rising temperature enhances the accessibility of clustered nitrogen-nitrogen defect interactions. Crucially, it is not the energetic contributions of nitrogen pairs alone that drive the transition---since such motifs exist even in dispersed configurations---but the point at which clustered configurations start to dominate the partition function due to their abundance compared to dispersed configurations.

\section{Conclusion}

We have presented a modification of the SAMPLE approach capable of generating defective two-dimensional structures. Using this approach, we have generated more than 100 million configurations of NDG, enabling us to explore the compositional phase diagram and the thermodynamic properties. In the phase diagram, we observe a sharp change in the convex hull gradient at $\theta$~=~0.25, corresponding to the stoichiometry C\textsubscript{3}N, which is in agreement with literature. The internal energy and entropy exhibit a sigmoidal behaviour when plotted against the temperature. The heat capacity shows a characteristic peak, which is the result of a combinatorially driven order-disorder transition. The positions of the peaks strongly depend on the chemical potentials of nitrogen and carbon, which affect the thermodynamic favourability of incorporating nitrogen into graphene. Future work could focus on extending this approach to other defective 2D materials, such as the full boron-nitrogen-carbon phase diagram and defects such as the Stone-Wales defect, to explore their configurational landscapes and phase behavior under varying thermodynamic conditions. More efficient data processing will enable the exploration of even larger configuration spaces.

\section*{Acknowledgments}

The authors acknowledge financial support from the UKRI Future Leaders Fellowship program (MR/X023109/1), an AIchemy Pump Prime grant (ID: 122408), funded through the EPSRC (EP/Y028759/1), and UKRI Horizon grants (ERC StG, EP/X014088/1 and MSCA-PF, EP/Y024923/1). We gratefully acknowledge Oliver T. Hofmann and Christoph Wachter for support with the SAMPLE code and Livia Bartók-Pártay and Thanawitch Chatbipho for many fruitful discussions. High-performance computing resources were provided via the EuroHPC supercomputer LUMI, hosted by CSC (Finland) and the LUMI consortium through a EuroHPC Regular Access call (EHPC-REG-2023R03-123). Additional computation resources were provided by the Scientific Computing Research Technology Platform of the University of Warwick, the EPSRC-funded Materials Chemistry Consortium (EP/R029431/1, EP/X035859/1), and the UK Car-Parrinello consortium (EP/X035891/1) for the ARCHER2 UK National Supercomputing Service, the EPSRC-funded HPC Midlands+ computing centre for access to Sulis (EP/P020232/1).

\section*{Data and Code Availability}

The data supporting this research is publicly available on the NOMAD repository: \href{https://doi.org/10.17172/NOMAD/2025.09.13-1}{https://doi.org/10.17172/NOMAD/2025.09.13-1}. The SAMPLE code is publicly available at \href{https://gitlab.tugraz.at/othgroup/sample}{https://gitlab.tugraz.at/othgroup/sample}.

\bibliography{references} 
\end{document}